\documentclass[onecolumn,10.9pt]{article}
\usepackage[top=1.0in, bottom=1.2in, left=1.2in, right=1.2in]{geometry}
\usepackage{amsmath,amsfonts,amscd,amssymb}
\usepackage{graphicx}
\usepackage{epstopdf}
\usepackage{overpic}
\usepackage{cancel}
\usepackage{rotating}
\usepackage{url}
\usepackage{caption}
\usepackage{color}
\usepackage{rotating}
\usepackage{multirow}
\usepackage{wrapfig}
\usepackage{mathtools}
\usepackage{subeqnarray}
\usepackage{setspace}
\usepackage{palatino} 
\usepackage{hyperref}
\usepackage[numbers,sort&compress]{natbib}

\usepackage{mathabx}

\usepackage{subcaption}

\usepackage[bottom,flushmargin,hang,multiple]{footmisc}
\usepackage{lipsum}
\newcommand\blfootnote[1]{%
  \begingroup
  \renewcommand\thefootnote{}\footnote{#1}%
  \addtocounter{footnote}{-1}%
  \endgroup
}

\definecolor{header1}{cmyk}{0,0,0,1}

\title{Data-Driven Modeling for Transonic Aeroelastic Analysis}

\author{\normalsize{Nicola Fonzi$^{1}$, Steven L. Brunton$^{2}$ and Urban Fasel$^{3}$}\\
\footnotesize{$^1$ Department of Aerospace Science and Technology, Politecnico di Milano, Milano, 20156, Italy\blfootnote{Corresponding author (nicola.fonzi@polimi.it).}}\\
\footnotesize{$^2$ Department of Mechanical Engineering, University of Washington, Seattle, WA 98195, United States}\\
\footnotesize{$^3$ Department of Aeronautics, Imperial College, London, England SW7 2AZ, United Kingdom}
}
\date{\today}
\vspace{-1.2in}
\begin{document}

\maketitle

\begin{abstract}
    Aeroelasticity in the transonic regime is challenging because of the strongly nonlinear phenomena involved in the formation of shock waves and flow separation. In this work, we introduce a computationally efficient framework for accurate transonic aeroelastic analysis. We use dynamic mode decomposition with control (DMDc) to extract surrogate models from high-fidelity computational fluid dynamics (CFD) simulations. Instead of identifying models of the full flow field or focusing on global performance indices, we directly predict the pressure distribution on the body surface. The learned surrogate models provide information about the system stability and can be used for control synthesis and response studies. Specific techniques are introduced to avoid spurious instabilities of the aerodynamic model. We use the high-fidelity CFD code SU2 to generate data and test our method on the benchmark super critical wing (BSCW). Our python-based software is fully open-source and will be included in the SU2 package to streamline the workflow from defining the high-fidelity aerodynamic model to creating a surrogate model for flutter analysis.
\end{abstract}

\section{Introduction}
\begin{figure}[t]
    \centering
    \includegraphics[width=0.8\textwidth]{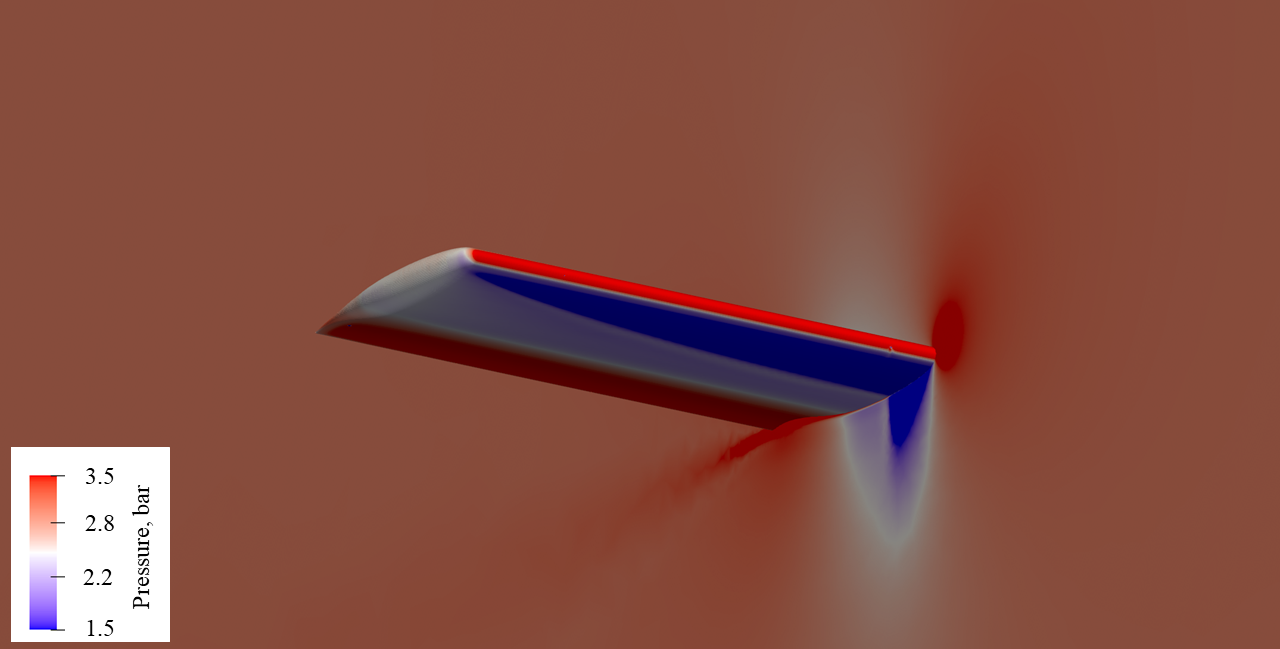}
    \vspace{5pt}
    \caption{Pressure contour of the Benchmark Super Critical Wing during a pitching down maneuver, highlighting the shock position on the pressure side of the wing.}
    \label{fig:mesh}
\end{figure}

Nonlinear aeroelasticity has been the focus of great interest in recent years. Several workshops have been organised on this research topic, including the recent NASA aeroelastic prediction workshops~\cite{heeg_overview_2013, heeg_plans_2015, noauthor_3rd_2021}. While several tools are available and recognized for certification in the subsonic (linear) regime~\cite{Theodorsen:35, albano_doublet-lattice_1969, dowell:1973, leishman:06}, flutter prediction in transonic (nonlinear) conditions still pose great challenges. In general, we can identify two main sources of nonlinearities: structural and aerodynamic. Structural nonlinearities can originate from either concentrated or distributed nonlinearities. The former is typically the case for freeplay (or cubic hardening) nonlinearities in the control surfaces, while the distributed nonlinearities usually originate from large structural displacements. One main objective of the third AIAA aeroelastic prediction workshop is to investigate nonlinear aeroelastic phenomena induced by large deflections, studied on the Pazy wing test case~\cite{drachinsky_flutter_2022, mertens_experimental_2022, righi2022uncertainties}. On the other hand, aerodynamic nonlinearities are a particular challenge for flutter prediction in the transonic regime. These nonlinearities can originate from separated flow caused by large angles of attack and shock waves. In the third aeroelastic prediction workshop, these phenomena are investigated by studying the benchmark super critical wing (BSCW) operating in the transonic regime and at relatively large angles of attack~\cite{poplingher_transonic_2022, poplingher2022flutter, poplingher2023stall}. 
%
The wing has a rectangular planform, a supercritical profile, and is fixed on one side to the wind tunnel wall, while the other side has a free tip. The wing can be set to have free pitch and plunge, or pitch only. The structure is rigid enough to only consider these two degrees of freedom. A snapshot of the wing pressure distribution during a pitching down maneuver is shown in Fig.~\ref{fig:mesh}. A shock wave can clearly be seen on the lower side of the wing. Depending on the operating condition, a shock wave and strong separation can be present on the suction side, or both on the suction and pressure side. At high angle of attack, stall flutter can be observed; a phenomena that can create instabilities due to the interaction of the wing structural modes, the shock wave, and the separation\cite{poplingher2023stall}. 

For both structural and aerodynamic nonlinearities, applying conventional, linear analysis methods is generally insufficient. 
Extensions exist to relax these linear assumptions in a number of sub-cases.  
However, fully accounting for all nonlinear phenomena requires new approaches, particularly for aerodynamic nonlinearities. The strong nonlinearities involved in the formation and destruction of shock waves can only be accurately captured with aerodynamic solutions obtained from high-fidelity CFD simulations. 
Thus, these numerical approaches are necessary if flutter in the transonic regime or buffeting is investigated; however, the computational cost of these high-fidelity simulations is often prohibitively expensive~\cite{heeg_overview_2016}, especially in early design stages.

To enable efficient analysis in these conditions, there is a need for surrogate models that accurately represent the nonlinearities at significantly lower computational cost. An extended review of fluid-structure interaction modeling is presented in \cite{dowell:2001}, where several reduced-order modeling methods are also presented. Recent data-driven modeling approaches are promising candidates that enable the identification of accurate and tractable models of nonlinear dynamical systems \cite{Kutz2016book}. Early work applied nonlinear systems theory~\cite{silva_application_1993} or modal-based techniques~\cite{Dowell:1997, hall2000proper} to develop reduced-order unsteady aerodynamic response models in the transonic regime. More recent work on data-driven transonic flutter prediction used the auto regressive moving average (ARMA) method in a multioutput scheme and was successfully applied to predict flutter of the BSCW~\cite{argaman2019multioutput}. 
The identified ARMA model is used to predict the evolution of global quantities (i.e., the structural modes) in time, and it is also used to define a stability criterion to predict the instability point. Recently, deep neural networks have also been used~\cite{li_deep_2019}, although this approach typically relies on large datasets and the resulting models may be difficult to interpret. In contrast, the sparse identification of nonlinear dynamics (SINDy) technique~\cite{brunton2016discovering} has been used to extract interpretable, minimal-order descriptions of oscillating shock wave dynamics on two-dimensional airfoils at transonic buffet conditions~\cite{sansica2022system}.

Several recent efforts have further extended research capabilities in the field of flutter prediction and reduced-order modeling (ROM), including: harmonic balance~\cite{simiriotis_flutter_2022}, linearized frequency domain~\cite{stanford_aeroelastic_2022}, and Volterra methods~\cite{brown_convolutionvolterra_2022}. The harmonic balance requires the reformulation of the fluid equations under the assumption of a pure harmonic motion~\cite{simiriotis_flutter_2022}. Recalling that at the flutter point the aeroelastic solution is purely harmonic, the aerodynamic forces and the structural solution can be iteratively solved until a null damping is found. At that point, the coupled equations are valid and the flutter point is found. 
The second work considers linearized frequency domain aerodynamics to study the flutter envelope of the Pazy wing~\cite{stanford_aeroelastic_2022}. First, a static nonlinear solution is computed at each operating condition and reduced frequency, and the eigenmodes of the deformed configurations are extracted. Then, the generalized aerodynamic forces at each operating condition are computed for each mode. This can improve the computational efficiency compared to a fully coupled simulation. Also, the usual $V$-$g$, $V$-$f$ plots can be obtained, and the results may be used for control synthesis and other classical analysis. 
Finally, it is possible to construct data-driven models using Volterra kernels~\cite{brown_convolutionvolterra_2022}, which is a nonlinear extension of the aerodynamic impulse response method~\cite{wagner:25}. In practice, several simulations are run, starting from an equilibrium point. In each simulation, a single mode or multiple modes are excited with an impulse of a certain amplitude. When multiple modes are excited, the order of the inputs and their delay is important. Then, response kernels are obtained from the measured responses, and convolved in the prediction step. The challenge with this method is to identify the optimal combination and amplitudes of inputs. Also, the training is performed at one single operating condition and the accuracy of the method may degrade when deviating from this condition. If the correct combination of inputs and the correct operating point can be identified, the limit cycle oscillation (LCO) amplitude and the linear flutter onset can be accurately predicted.

A promising approach that provides physically interpretable models is the dynamic mode decomposition with control (DMDc)~\cite{proctor_dynamic_2016}, previously applied by the authors to a flexible wing for real-time prediction and control of nonlinear unsteady aeroelastic responses~\cite{fonzi_data-driven_2020}. DMDc is an extension of the dynamic mode decomposition (DMD), which enables the discovery of dynamical systems from high-dimensional data by decomposing complex dynamics into simple representations based on spatio-temporal coherent structures~\cite{rowley_spectral_2009, schmid_dynamic_2010, tu_dynamic_2014, williams_datadriven_2015,Kutz2016book,askham_variable_2018}. It is strongly connected to the Koopman operator~\cite{Mezic2004physicad,Mezic2005nd,rowley_spectral_2009} and is therefore a promising candidate to model the nonlinear dynamics inherent to transonic flutter. Moreover, compared to black-box machine learning methods, DMD/DMDc models are physically interpretable and can be identified with limited data.
Recently, DMD was applied to model unsteady flows and to analyse fluid–structure stability~\cite{yao_data-driven_2022}. The aerodynamic ROM considered as states the pressure coefficients in the entire bi-dimensional flowfield, and integrated them on the surface of a NACA0012 to obtain structural coupling. The training data was obtained using a high--fidelity CFD solution. The aerodynamics are still in the linear regime, allowing the model to be trained at one operating condition only, and scaling the aerodynamic forces with the dynamic pressure. The results are a promising application of DMDc to aeroelastic problems. 

Starting from these latest results, we present a complex three dimensional aeroelastic modeling scenario and methodology, leveraging the use of data-driven methods for aeroelastic problems.

\subsection{Main objectives}
In this work, we present a comprehensive framework that uses DMDc to create surrogate models of nonlinear aerodynamics for aeroelastic applications. Our methodology provides a fast and effective way of extracting ROMs from high-fidelity CFD simulations. The identified models provide information about the system stability that is useful for flutter analysis, and can also be used for control synthesis and response studies. The ROMs are represented by dynamical systems in state space, which are well known in the aeroelastic community and can be easily included in other design procedures. DMDc is designed to work in the low data limit, reducing the number of high-fidelity simulations required for training. 
DMDc has been widely applied to discover coherent spatio-temporal structures of fluid flows. In our work, inspired by well established wing aerodynamics panel codes, we reduce the aerodynamic problem to the surface of the wing and identify coherent spatio-temporal pressure distributions, as opposed to full flow fields. We introduced this idea in~\cite{fonzi_data-driven_2020} and now extend it to aerodynamics in the transonic regime.

This approach can drastically reduce the computational complexity of the problem compared to considering the entire three-dimensional flow field. 
To further reduce the rank of the aerodynamic model, a stabilization procedure is used, based on ideas introduced in~\cite{brunton2013reduced, brunton2014state, Hemati2016aiaa,hickner2022data}. Finally, the entire set of methods is implemented in an open--source python framework. The framework allows one to automatically create ROMs from high fidelity CFD training data and to use the ROMs for aeroelastic analyses.
We test our code on the BSCW~\cite{raveh2018analyses}, for which a large amount of data is available for comparison, including wind tunnel test data. The BSCW wing was recently used to study bidimensional aerodynamic reduced-order models for flutter prediction and uncertainty quantification~\cite{righi2021uncertainties,righi2022rom}, which makes the wing a good platform for the scope of our research. 
We use the high-fidelity open-source code SU2 to generate the aerodynamic data~\cite{economon_su2_2016}. In a future step, we will include our software directly in the SU2 package, to streamline the workflow from the definition of the high-fidelity aerodynamic model to the creation of the surrogate model.

\section{Methods}
We use DMDc to generate surrogate aerodynamic models from CFD data. DMDc has been used previously for aeroelastic problems to generate monolithic models (i.e., state space systems that incorporate both the aerodynamics and the structural dynamics) purely from data~\cite{fonzi_data-driven_2020}. In the context of this research, due to the strong nonlinear aerodynamics, we treat the structural modeling and the aerodynamic modeling separately. First, we generate data using the SU2 open-source CFD code. We use a custom python framework in SU2 to manage the aeroelastic simulation~\cite{fonzi2022extended}. The structural system is modeled via a linear combination of structural modes, while the aerodynamic model is based on the Reynolds averaged Navier-Stokes (RANS) equations. The structural displacements can either be imposed, or computed via a generalized-alpha method. This enables aerodynamic response analyses with imposed motion, and fully coupled aeroelastic simulations. The rest of this section is divided in three parts, following the different novelties introduced in the work.

\subsection{Surface DMDc in transonic application}

To generate a DMDc model, we first specify the inputs and the states of the model. The goal is to fit the evolution of the states of a dynamic system as:
\begin{equation}
    \mathbf{x}^{k+1} = \mathbf{A}\mathbf{x}^k + \mathbf{B}\mathbf{u}^k
\end{equation}
where $\mathbf{x}$ is the state vector, $\mathbf{u}$ the input vector, and $k$ is the time index. $\mathbf{A}$ and $\mathbf{B}$ are the unknown matrices. Typical applications of DMD in the field of aerodynamics use the entire flowfield as a state. In this research, inspired by previous work on panel methods, we reduce the tridimensional problem to an equivalent bidimensional one, by only using the pressure distribution on the surface as a state. The tridimensional behaviour will then be implicitly modeled by the DMD modes. As for the input, we use the modal amplitudes and their first and second derivatives. Another option, as shown in~\cite{yao_data-driven_2022}, would be to time-embed the modal amplitudes directly. That is, time-embedding the modal amplitudes signifies that past amplitudes are taken into account, not only the ones at time $k$. However, this approach is not considered here, because in the initialisation phase the past modal amplitudes are not known and extrapolation would be required. On the other hand, first and second derivatives are always known, either from the structural integrator or from inputs provided by the user. 

The second step is to collect data by computing a full simulation imposing inputs to the aerodynamics. There is no requirement about the shape of the inputs, nor the amplitude. However, it is important that all the inputs are activated. In general, it is a good practice to use modal amplitudes that have a similar effect on the aerodynamics (e.g.measured via macroscopic quantities such as the lift). The state and input trajectories of the full simulation are stored in three matrices. The first two gather the evolution of the states:
\begin{equation}
    \mathbf{X}=[\mathbf{x}^1,\mathbf{x}^2,\mathbf{x}^3, \dots ,\mathbf{x}^{n-1} ],\ \ \ \ X'=[\mathbf{x}^2,\mathbf{x}^3,\mathbf{x}^4, \dots ,\mathbf{x}^{n} ]
\end{equation}
where $n$ signifies the total simulation length in terms of number of iterations. In this application, the state vector will be:
\begin{equation}
    \mathbf{x} = \begin{bmatrix}
        Cp_1, & 
        Cp_2, & 
        Cp_3, & 
        \dots, &
        Cp_m
    \end{bmatrix}^T
\end{equation}
with $Cp$ the pressure coefficient, and the subscript identifying the index of a mesh vertex on the surface mesh of the body.

The last matrix registers the inputs:
\begin{equation}
    \boldsymbol{\Upsilon} =
    \begin{bmatrix}
        \begin{bmatrix}
            u_1^1 & u_2^1 & \dots & u_l^1 \\
            u_1^2 & u_2^2 & \dots & u_l^2 \\
            \vdots & \vdots & \ddots & \vdots \\
            u_1^n & u_2^n & \dots & u_l^n \\
        \end{bmatrix}
        \begin{bmatrix}
            \dot{u}_1^1 & \dot{u}_2^1 & \dots & \dot{u}_l^1 \\
            \dot{u}_1^2 & \dot{u}_2^2 & \dots & \dot{u}_l^2 \\
            \vdots & \vdots & \ddots & \vdots \\
            \dot{u}_1^n & \dot{u}_2^n & \dots & \dot{u}_l^n \\
        \end{bmatrix}
        \begin{bmatrix}
            \ddot{u}_1^1 & \ddot{u}_2^1 & \dots & \ddot{u}_l^1 \\
            \ddot{u}_1^2 & \ddot{u}_2^2 & \dots & \ddot{u}_l^2 \\
            \vdots & \vdots & \ddots & \vdots \\
            \ddot{u}_1^n & \ddot{u}_2^n & \ddots & \ddot{u}_l^n \\
        \end{bmatrix}
   \end{bmatrix}^T
\end{equation}
with $l$ being the number of modes of the structural system. The different sets of inputs are the modal amplitudes, velocities, and accelerations.

The $\mathbf{X}$, $\mathbf{X}'$, and $\boldsymbol{\Upsilon}$ matrices are called the snapshot matrices. A first important analysis is the decomposition of the snapshot matrix $\mathbf{X}$ to extract the proper orthogonal decomposition (POD) modes. These modes are then used to project the states of the dynamical system to obtain the DMD modes. In the rest of the paper, we will call the DMD modes ``aerodynamic modes", as they contain the main spatio-temporal features of the aerodynamic system. The decomposition is performed using the singular value decomposition (SVD). This reads:
\begin{equation}
    \mathbf{X} = \mathbf{U}\mathbf{S}\mathbf{V}^{*}
\end{equation}
where $\mathbf{U}$ is the matrix containing the POD modes, also interpreted as the spatial coherent structures in the snapshots, $\mathbf{S}$ is a diagonal matrix containing the non-negative singular values, in decreasing order, that identifies the variance of the modes, and $\mathbf{V}$ is the matrix containing the orthogonal right singular vectors. The symbol $^{*}$ signifies the complex conjugate transpose operator. The POD modes are of great interest, as the aerodynamics can be projected onto the subspace spanned by these modes, similar to the linear modal analysis of a structural element. In the results section, we will show that they identify the most important features of the aerodynamic system. Further, they can be truncated, retaining only the most important modes, as measured by their variance. This can be performed using the Frobenius norm, which is expressed for $\mathbf{X}$ as:
\begin{equation}
    ||\mathbf{X}||_F = \sqrt{\sum_{i=1}^{n-1} \mathbf{S}_{ii}}
\end{equation}
with $\mathbf{S}_{ii}$ identifying the diagonal element of the $\mathbf{S}$ matrix. When truncating the SVD matrices, the root mean square error $\epsilon$ will result in:
\begin{equation}
    \epsilon = \frac{||\mathbf{X}-\mathbf{X}_r||_F}{||\mathbf{X}||_F} = \sqrt{\frac{\sum_{i=r+1}^{n-1} \mathbf{S}_{ii}}{\sum_{i=1}^{n-1} \mathbf{S}_{ii}}}
\end{equation}
with $r$ being the truncation point. Thus, the singular values alone are enough to decide where to truncate the SVD, for a given accepted error in the reconstruction of the matrix $\mathbf{X}$.

The third step is to generate a surrogate model through the DMDc procedure~\cite{proctor_dynamic_2016}, by first stacking the snapshot matrices together:
\begin{equation}
    \boldsymbol{\Psi}=[\mathbf{X};\boldsymbol{\Upsilon}]
\end{equation}
and then imposing the following relation:
\begin{equation}
    \mathbf{X}'=[\mathbf{A},\mathbf{B}]\boldsymbol{\Psi}
\end{equation}
where the notation $[\mathbf{X};\boldsymbol{\Upsilon}]$ signifies the vertical stacking of matrices and $[\mathbf{A},\mathbf{B}]$ the horizontal stacking.

We can solve for the A and B matrices using a Moore-Penrose pseudoinverse of the matrix $\Psi$, based again on the SVD:
\begin{equation}
    \boldsymbol{\Psi} = \bar{\mathbf{U}}\bar{\mathbf{S}}\bar{\mathbf{V}}^{*}.
\end{equation}
The inversion of this matrix then results in:
\begin{equation}
    [\mathbf{A},\mathbf{B}] = \mathbf{X}'\bar{\mathbf{V}}\bar{\mathbf{S}}^{-1}\bar{\mathbf{U}}^{*}
\end{equation}
as the $\bar{\mathbf{S}}$ matrix is diagonal and is easily inverted. 
The resulting dynamical system is then projected onto the truncated POD modes to reduce the size of the system:
\begin{equation}
    \tilde{\mathbf{A}} = \mathbf{U}_r^{*} \mathbf{A} \mathbf{U}_r
\end{equation}
and
\begin{equation}
    \tilde{\mathbf{B}} = \mathbf{U}_r^{*} \mathbf{B}.
\end{equation}

Starting from arbitrary initial conditions, the reduced-order state--space system can then be evolved in time:
\begin{equation}
    \mathbf{\tilde{x}}_{k+1} = \tilde{\mathbf{A}}\mathbf{\tilde{x}}_k + \tilde{\mathbf{B}}\mathbf{u}_k
\end{equation}
and the full aerodynamic state can be recovered as: 
\begin{equation}
    \mathbf{x} = \mathbf{U}_r \mathbf{\tilde{x}}.
    \label{eq:recoverState}
\end{equation}

The DMD modes are then obtained by diagonalizing $\tilde{\mathbf{A}}$, as presented in~\cite{proctor_dynamic_2016}.

\subsection{Stabilization procedure}

The state--space dynamical system obtained using DMDc can be slightly unstable, depending on the truncation point and the input signal that is used. This is a mathematical artifact and it is necessary to stabilize the dynamical system to improve the prediction accuracy of the ROM. 
The stabilization approach is inspired by recent works on reduced-order unsteady aerodynamic and aeroelastic models~\cite{brunton2013reduced, brunton2014state, hickner2022data}. In their context, the goal was to obtain physically interpretable ROMs, as obtained via the eigensystem realisation algorithm (ERA)~\cite{Juang1985jgcd,Ma2011tcfd}. In our work, we use their approach to create better conditioned DMDc models to improve the stability. The core idea is to distinguish between the unsteady part, that has to be modeled via a dynamic model, and the quasi-steady part. It must be noted that the following method can only be applied if the training signal and the aerodynamics respect two important assumptions:
\begin{itemize}
    \item The aerodynamic response must be symmetric with respect to the inputs.
    \item The training signal must allow the aerodynamics to reach the steady state condition.
\end{itemize}

The first requirement is quite stringent, as it imposes that a positive input in a mode has the exact opposite effect of a negative input, of same amplitude. This is not always the case, for example a shock wave in a cambered profile does not develop in the same way on the pressure side and the suction side when it pitches. However, the amplitude of the mode can be set so that the aerodynamic response can be accurately linearized.

\begin{figure}[t]
    \centering
    \includegraphics[width=\textwidth]{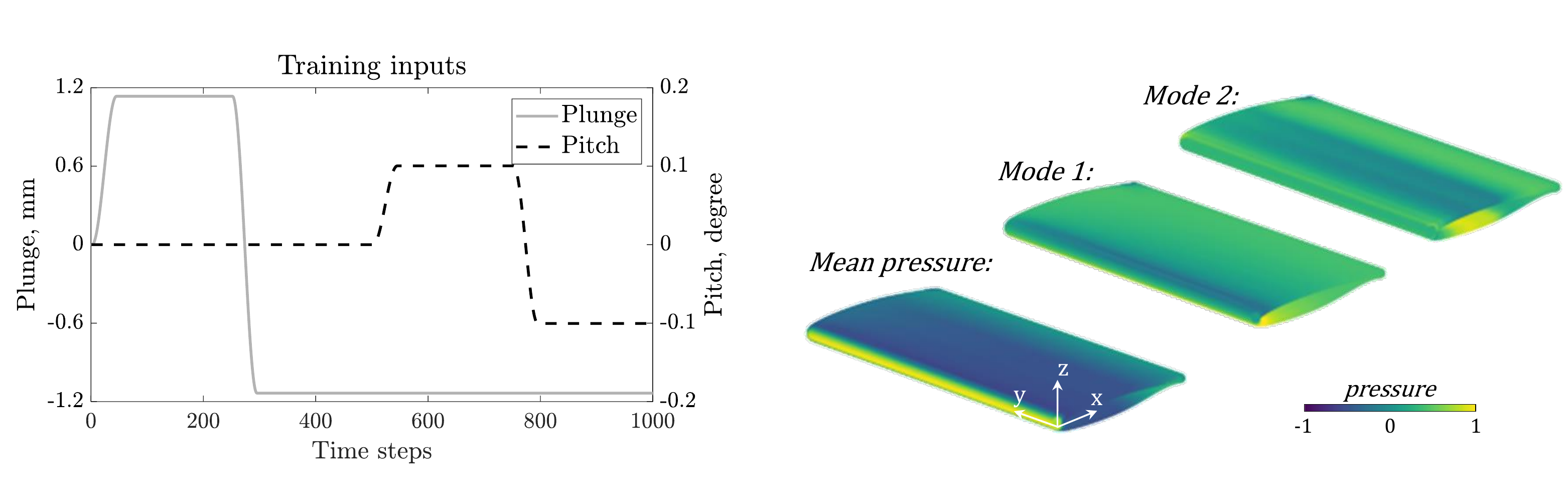}
    \caption{(Left) Plunge and pitch training input. (Right) Wing normalized pressure distributions (mean, mode 1, and mode 2).}
    \label{fig:tutorial1}
\end{figure}

The second requirement can be met by choosing an appropriate training signal. In practice, we choose a time window that is long enough for the aerodynamics to reach steady state where the inputs are not changing. A best practice is to change the inputs one at a time to capture the steady effects separately. An example of an appropriate training signal, for two modes only, is reported in Fig.~\ref{fig:tutorial1} (left). Different training signals have been successfully applied in other research, including Walsh functions~\cite{silva_simultaneous_2008}, blended impulses~\cite{goto2022data} and random inputs. However, for the reasons explained above, they have not been used in the present research.

The steady effect of the inputs can then be defined as follows. First, we select characteristic snapshots where the aerodynamics are in steady state. We select the initial snapshot and one additional snapshot per mode. The characteristic snapshots are called $\mathbf{x}_{{SS}_i}$, where $i$ is the mode index ($0$ for the initial condition, and $1, 2, ...$ for the structural mode number for the other snapshots). In the example shown in Fig.~\ref{fig:tutorial1}, these snapshots would be taken at time iteration 0, 450, and 1000. After selecting characteristic snapshots, we compute the slope of the quasi-steady aerodynamics for each mode, as:
\begin{equation}
    \frac{\partial \mathbf{x}}{\partial u_i} = (\mathbf{x}_{{SS}_i} - \mathbf{x}_{{SS}_{i-1}})/u_i.
\end{equation}
This steady part can then be subtracted from the aerodynamic state as:
\begin{equation}
    \mathbf{x}_{\text{unsteady}} = \mathbf{x} - \sum_{i=1}^{l} \frac{\partial \mathbf{x}}{\partial u_i} u_i,
\end{equation}
and we obtain a new snapshot matrix by stacking the snapshots of the unsteady part $X_{\text{unsteady}}$. The new input snapshot matrix is also modified by only including the derivatives of the input signals $\Upsilon_{\text{unsteady}}$. The standard DMDc algorithm can then be performed on these snapshot matrices and we obtain a model that represents the unsteady-only behaviour of the aerodynamics. 

It is important to note that if this procedure is used, the full state reconstruction is not obtained as in equation \ref{eq:recoverState}, but rather by:
\begin{equation}
    \mathbf{x} = \mathbf{U}_r\tilde{\mathbf{x}}_{\text{unsteady}} + \sum_{i=1}^{l} \frac{\partial \mathbf{x}}{\partial u_i} u_i.
\end{equation}

\subsection{Fully coupled aeroelastic ROM}

The procedure described in the previous sections is used when an aerodynamic prediction is sought. That is, a set of structural inputs are known and provided to the reduced-order model, and the aerodynamic forces are obtained as a result. However, we are often interested in obtaining a fully coupled simulation, where, once the aerodynamic model is trained, we can integrate it into an aeroelastic simulation. Moreover, this can further be extended by including a control law. In this case, the prediction coming from the DMD-ROM is directly used as an input for a structural model. The latter equations read:
\begin{equation}
    \mathbf{M}\ddot{\mathbf{x}}_s + \mathbf{C}\dot{\mathbf{x}}_s + \mathbf{K}\mathbf{x}_s = \boldsymbol{\Xi}\mathbf{F}_a
\end{equation}
where $\mathbf{x}_s$ is the vector of structural degrees of freedom (displacements in the three spatial directions), $M$, $C$, and $K$ are respectively the mass, damping, and stiffness matrices, and $\mathbf{F}_a$ is the vector containing the aerodynamic forces. $\Xi$ is an interpolation matrix, this can be constructed using different kernels. In our case, we use a radial basis function (RBF) kernel CP-C2 that recovers rigid translations~\cite{de_boer_mesh_2007}.

The aerodynamic forces are computed from the aerodynamic state $\mathbf{x}$, multiplying the pressure coefficients $C_p$ (contained in $\mathbf{x}$) with the surface area of the mesh cells $S$, and the matrix containing the normals of these cells $N$. In practice, the structural equations are further projected onto a set of structural modes $\Phi$, obtained via a generalized eigenvalue analysis:
\begin{equation}
    \boldsymbol{\Phi}^T\mathbf{M}\boldsymbol{\Phi}\ddot{\mathbf{x}}_s + \boldsymbol{\Phi}^T\mathbf{C}\boldsymbol{\Phi}\dot{\mathbf{x}}_s + \boldsymbol{\Phi}^T\mathbf{K}\boldsymbol{\Phi}\mathbf{x}_s = \boldsymbol{\Phi}^T\boldsymbol{\Xi} \mathbf{S}\mathbf{N}\mathbf{x}.
\end{equation}

The general tool described in this section is valid for flexible structures with an arbitrary number of modes, whereas for the BSCW considered in this work, we use two modes only.

\subsection{Open--Source Framework}

\begin{figure}
    \centering
    \includegraphics[width=\textwidth]{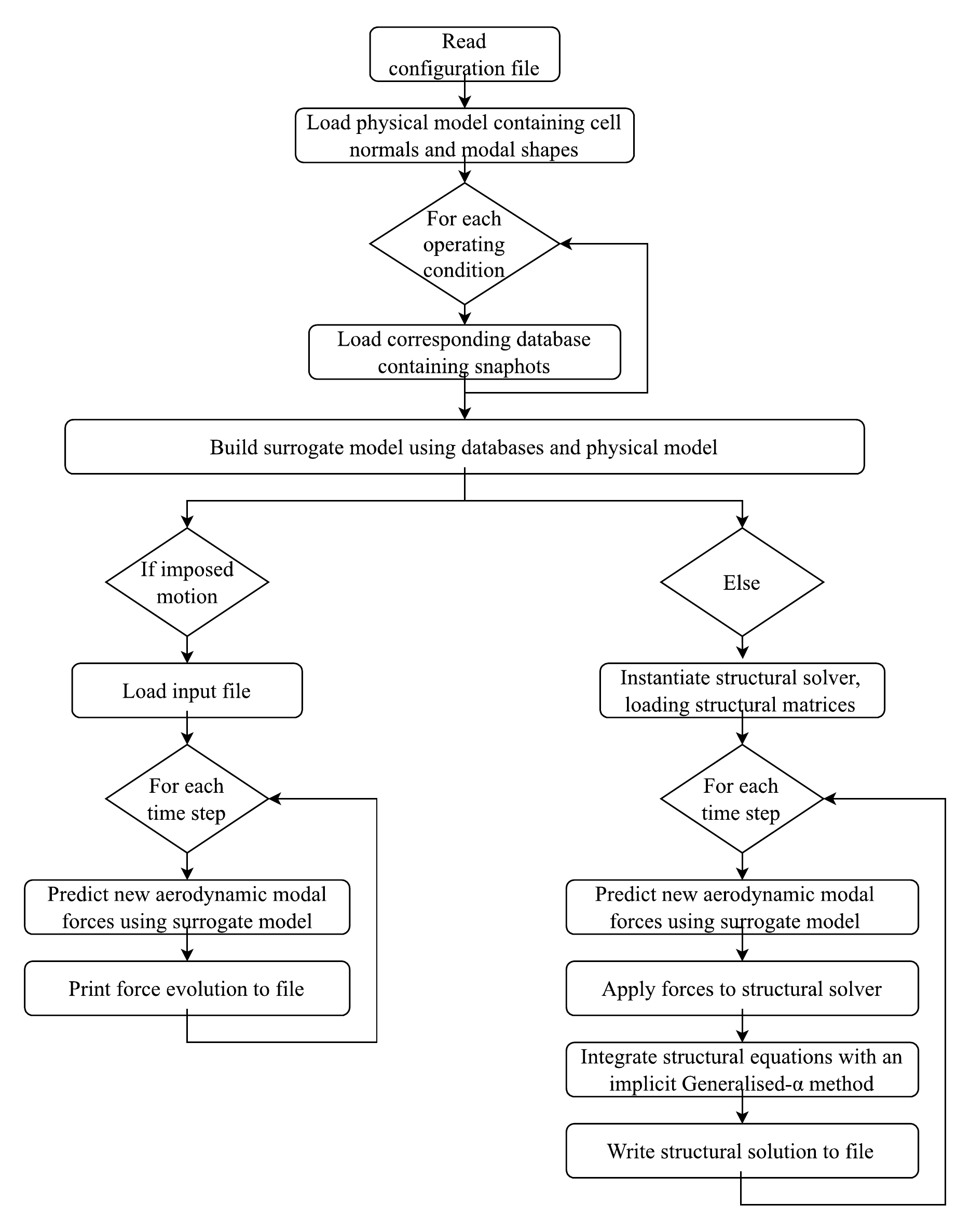}
    \caption{Flowchart of the framework. The code can be found on: \href{https://github.com/Nicola-Fonzi/pysu2DMD}{https://github.com/Nicola-Fonzi/pysu2DMD}}
    \label{fig:flowchart}
\end{figure}

The entire set of methods developed in this work has been included in a comprehensive framework, designed to natively work with the CFD solver SU2. The program is implemented in Python, for ease of portability and user--friendliness. To maximise the performance of the simulation, the transfer of information is directly performed using the shared memory. Following the typical workflow in a SU2 analysis, the software is controlled via the use of a configuration file, where specific keywords activate the various parts of the code. 
The flowchart representing the framework is reported in Fig.~\ref{fig:flowchart}.

It can be seen that two main possibilities are available: we can either use the driver to compute the aerodynamic forces, given the modal history, or we can actually compute a fully coupled aeroelastic response. In the latter case, we effectively substitute the SU2 python package, which provides high fidelity aeroelastic simulations, with our new framework. The model is trained by reading the SU2 output files that store the pressure distribution on the surface of the body. In order to compute the forces starting from the pressure distribution, another output file from SU2 is read, where the positions of the vertices and their normals are saved. Once trained, the model can be reused to provide other simulations at much lower cost. The structural solution is computed via a Generalised-alpha integration in time, the very same used in the SU2 package. Also, the structural model that has to be loaded by the driver is in the same format as the one used by the SU2 python package. Therefore, full compatibility is offered between the two software packages.

The framework can be used to compute the aerodynamic forces given different input signals than the ones used for training (at the same operating condition), or it can be used to compute the aeroelastic response in that condition. 
So far, only one operating condition can be taken into account, and the snapshot matrices are stored in a single database.
In the future, one database per operating condition can be built. These databases can then be interpolated to create a linear parameter varying (LPV) model with recently proposed approaches~\cite{Hemati2016aiaa,fonzi_data-driven_2020,iannelli2021balanced,goizueta_fast_nodate}.

\begin{figure}[t]
    \centering
    \includegraphics[width=0.58\textwidth]{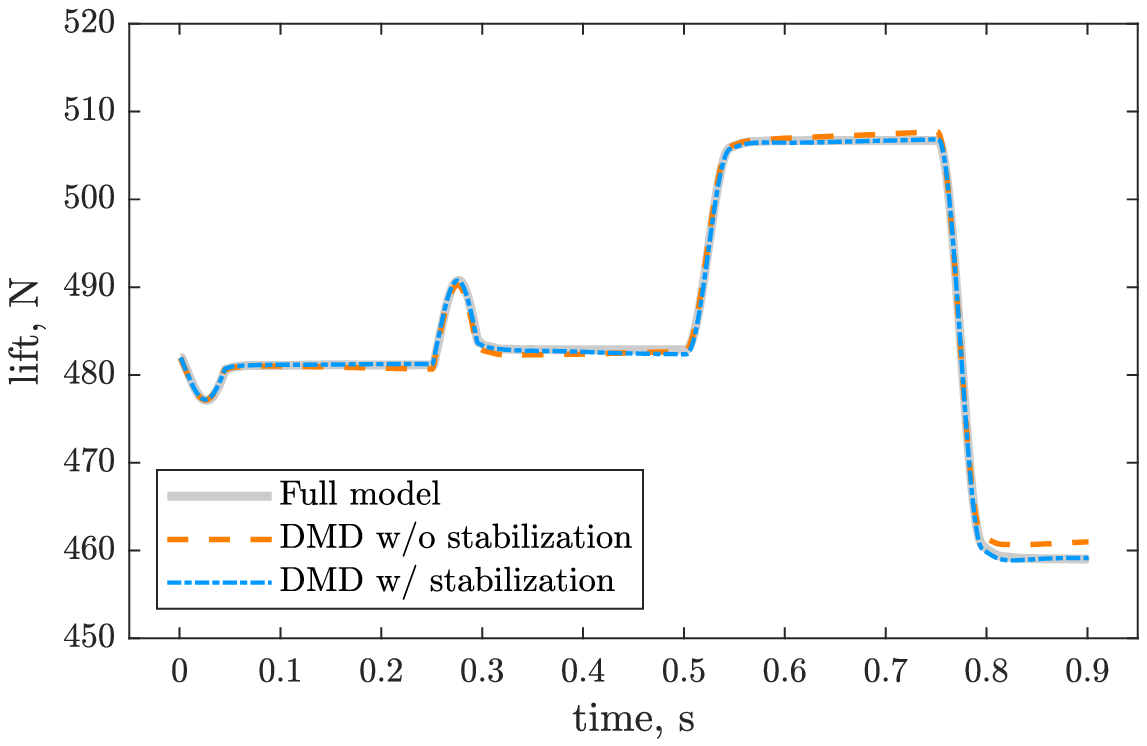}
    \hfill
    \includegraphics[width=0.37\textwidth]{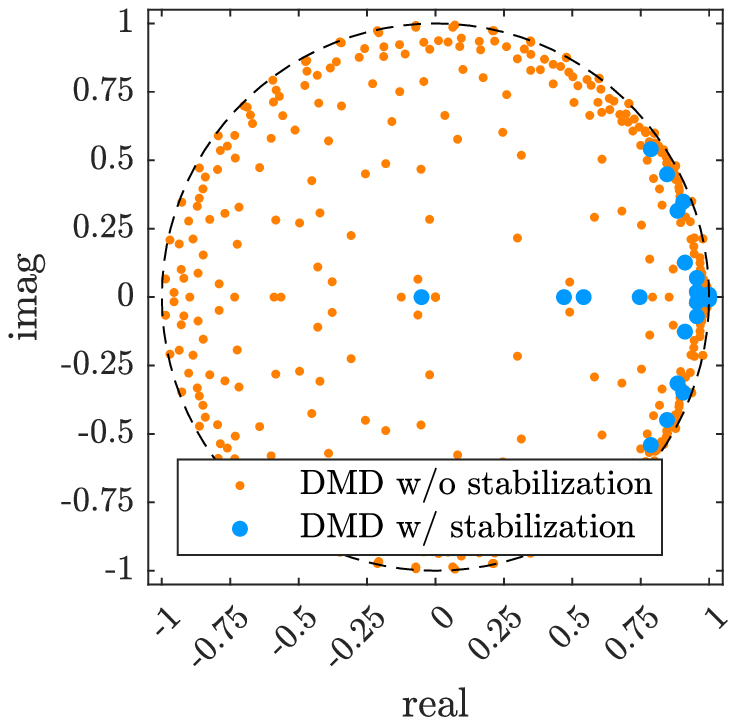}
    \caption{(Left) Lift force obtained via the high--fidelity SU2 python package and the DMD-ROM with and without stabilization. (Right) Eigenvalues of the DMD-ROM with and without the stabilization procedure.}
    \label{fig:LiftAndEV}
\end{figure}

\section{Results}

The main results of our work are summarized in a series of tutorials presented on \href{https://github.com/Nicola-Fonzi/pysu2DMD}{github}\footnote{https://github.com/Nicola-Fonzi/pysu2DMD}. The tutorials introduce the framework and showcase the possible applications of the methods and the software. In this section, we outline the challenges with increasing complexity and report the main results. For the following discussion we consider the BSCW, operating at Mach 0.74 and zero angle of attack, as explained in the second aeroelastic prediction workshop~\cite{heeg_overview_2013}. The wing is not operating in air, so care must be taken when setting the correct aerodynamic conditions. The physical model is suspended in the wind tunnel by springs that allow pitching and plunging with respect to the half chord. The training signal used is reported in Fig.~\ref{fig:tutorial1}.

\subsection{Aerodynamic modes}

In this first section, we build the database (snapshot matrices) and extract the POD modes of the pressure distribution. Indeed, the great advantage of DMDc is the formulation of the models in terms of physically interpretable modes, enabling for example the visualization of their evolution in time. No further manipulation of the data is done here.

In Fig.~\ref{fig:tutorial1}, the structural inputs (plunge and pitch) are shown on the left. On the right, we show the mean pressure distribution and the first two modes obtained from the snapshots of the system response to the modal amplitude input trajectories. The mean shows an almost uniform distribution, with stagnation at the leading edge. The first and second modes both show the decrease in pressure at the suction side for a positive pitch up motion. However, while the first mode might be representative also of a subsonic application, the second mode is characteristic of the transonic regime and the supercritical airfoil. On the pressure side, a strong and localised change in pressure can be observed, that will later form the shock seen in Fig.~\ref{fig:mesh}. The sign of the mode is changing in time, thus if the second mode has a negative sign in the simulation, the pressure will strongly decrease. On the suction side, the effect of the increased flatness of the profile is noticeable, with the flow accelerating twice, before recovering the pressure at the trailing edge. This is due to the fact that the maximum thickness point is located more towards the trailing edge with respect to subsonic profiles. At higher Mach number, this effect will also create a shock.

In this particular case, very few modes are required to reconstruct the entire pressure distribution. In a more complex aerodynamic condition, where strong separation is present, it is expected that more modes will be required, although the same principles hold.

\begin{figure}[t]
    \centering
    \includegraphics[width=\textwidth]{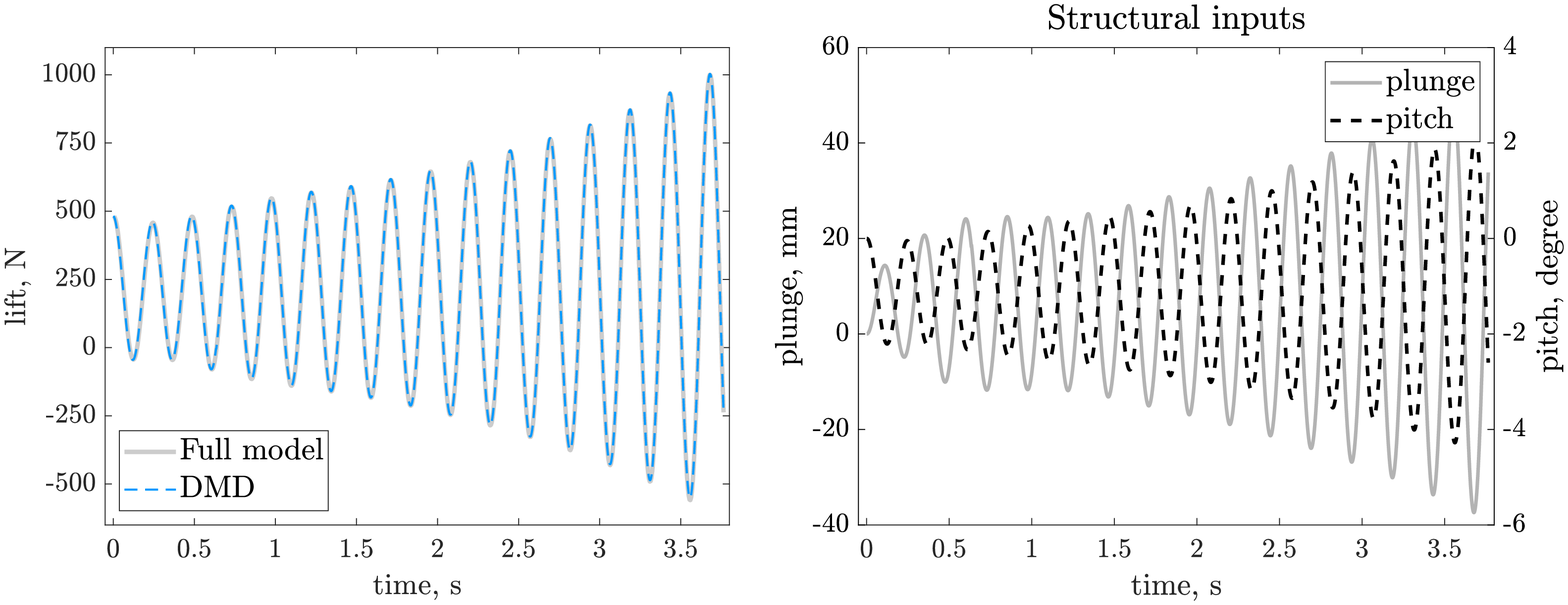}
    \caption{Lift force obtained via the high--fidelity SU2 python package and the DMD-ROM with stabilization, in the testing phase.}
    \label{fig:tutorial4}
\end{figure}

\subsection{Reproduction of the training signal}

An initial test is performed by comparing the ROM to the full order model aerodynamic response using the training input signal. The signal in Fig.~\ref{fig:tutorial1} is provided as an input to the ROM, and the aerodynamic forces are computed and compared with the full model.

The lift for both SU2 and the DMD-ROM is reported in Fig.~\ref{fig:LiftAndEV}. Here, we can see that a good match is obtained. However, a small drift is seen towards the end. This is due to the fact that a slightly unstable mode is present. This is nonphysical and can be corrected with the stabilization procedure presented in the previous section. Results after the stabilization are also reported in Fig.~\ref{fig:LiftAndEV}.

The stability of the system can also be assessed by plotting the eigenvalues of the $\tilde{A}$ matrix. If we recall that the dynamical system is expressed in state space and in discrete time, the eigenvalues, complex in general, should all lie within or on the unit circle. This is verified, as shown in Fig.~\ref{fig:LiftAndEV} on the right.

\subsection{Testing with a different input signal}

Next, we add complexly by trying to predict aerodynamic forces using the DMD-ROM for an input signal that is different from the one used for training. The input signal that is used is the structural history obtained from a fully coupled high--fidelity aeroelastic simulation at that operating condition. The system is dynamically unstable and produces an oscillation with a frequency of oscillation of around 4 Hz. The structural signal is thus substantially different from the blended step used for training.

The results in terms of lift force are compared for the full simulation and the ROM in Fig.~\ref{fig:tutorial4}.
An excellent match is seen also in this case, confirming the wide range of possible input signals that the DMD-ROM is able to model.

\subsection{Fully coupled aeroelastic prediction}

Finally, we couple our ROM with a structural model to predict the aeroelastic behaviour at that operating condition. We thus want to predict a complex model, where the linear aeroelastic system is unstable and an oscillation is developed with a certain frequency. This is an important step, indicating if our DMD-ROM can be used to predict the linear flutter boundary.

In the aerodynamic conditions used until now, the aeroelastic system is slightly past the flutter point. Indeed, we expect to see the appearance of an LCO. The flutter mechanism is dominated by the pitch, meaning that we expect to see large oscillations of the angle of attack, and smaller changes in the plunge amplitude. 

\begin{figure}[t]
    \centering
    \includegraphics[width=0.49\textwidth]{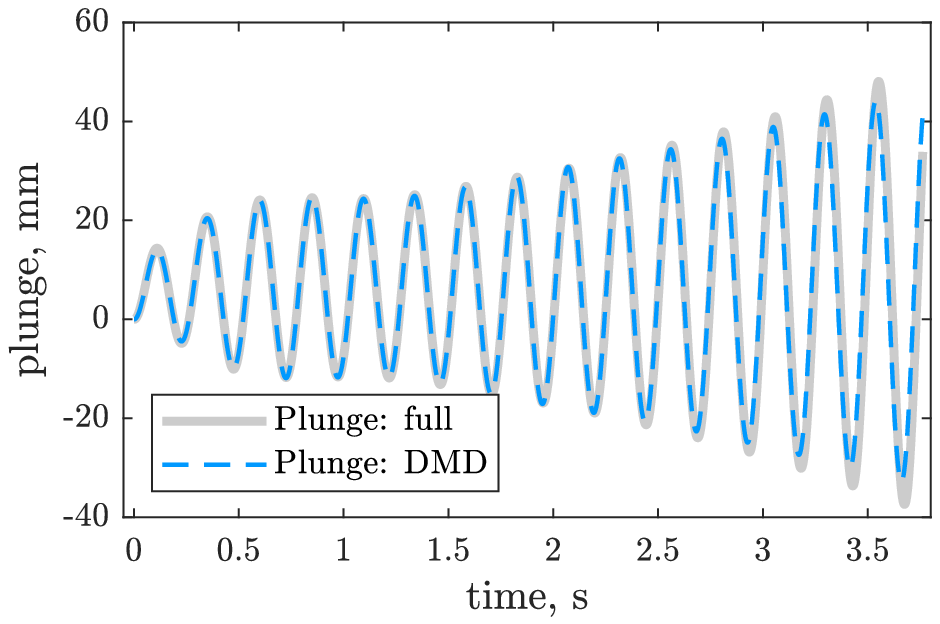}
    \hfill
    \includegraphics[width=0.49\textwidth]{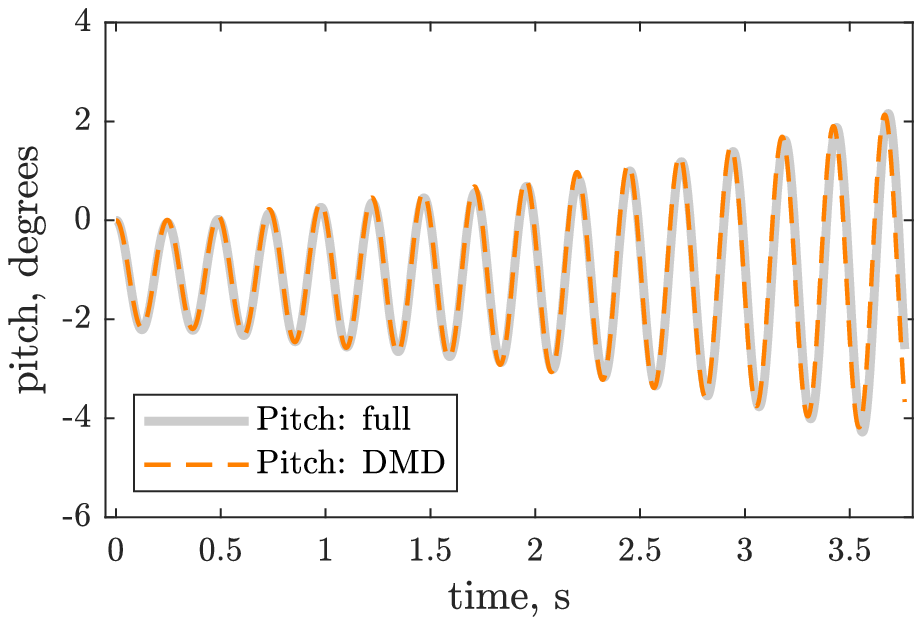}
    \caption{Plunge and pitch mode for the fully coupled simulation with SU2 and with the rank-30 DMD-ROM.}
    \label{fig:tutorial5}
\end{figure}

Figure~\ref{fig:tutorial5} shows the results of a time integration of the fully coupled aeroelastic simulation.
There is excellent agreement for the first second of the simulation (with time step 1 ms). A small discrepancy appears after 2.5s simulation time, due to the higher damping in plunge predicted via the reduced-order model. However, the instability is accurately predicted and flutter is identified. It is important to note that while the full aeroelastic simulation has millions of degrees of freedom, the reduced-order model only has 30 modes. This massive reduction in the problem dimension reduces the simulation time from 180 hours on a 40--core node of a cluster to some minutes on a single core. Further, the time integration of the ROM is not required to assess stability. The stability is assessed with an eigenvalue analysis that is performed in real time.

\begin{figure}[t]
    \centering
    \includegraphics[width=0.47\textwidth]{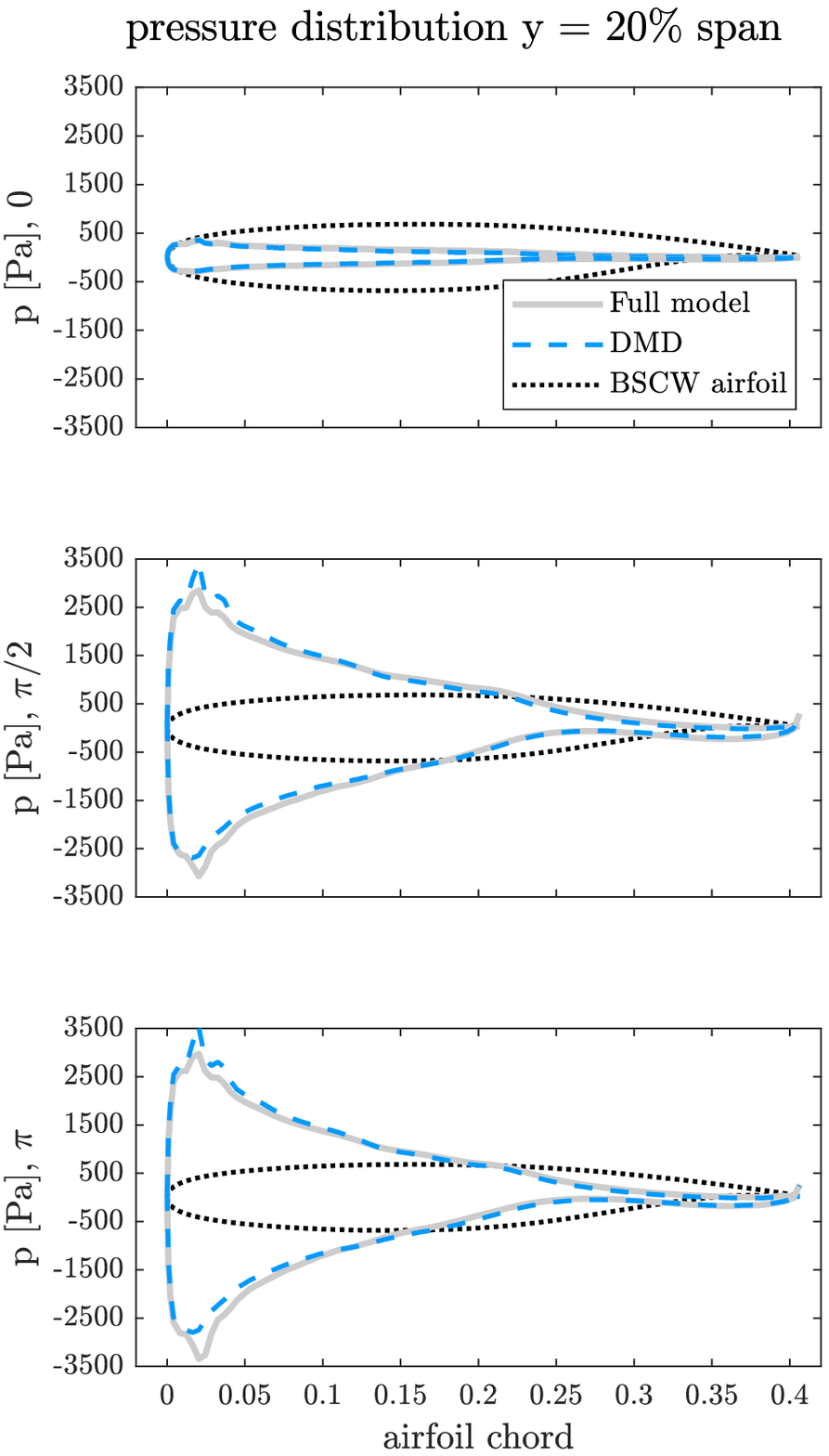}
    \hfill
    \includegraphics[width=0.47\textwidth]{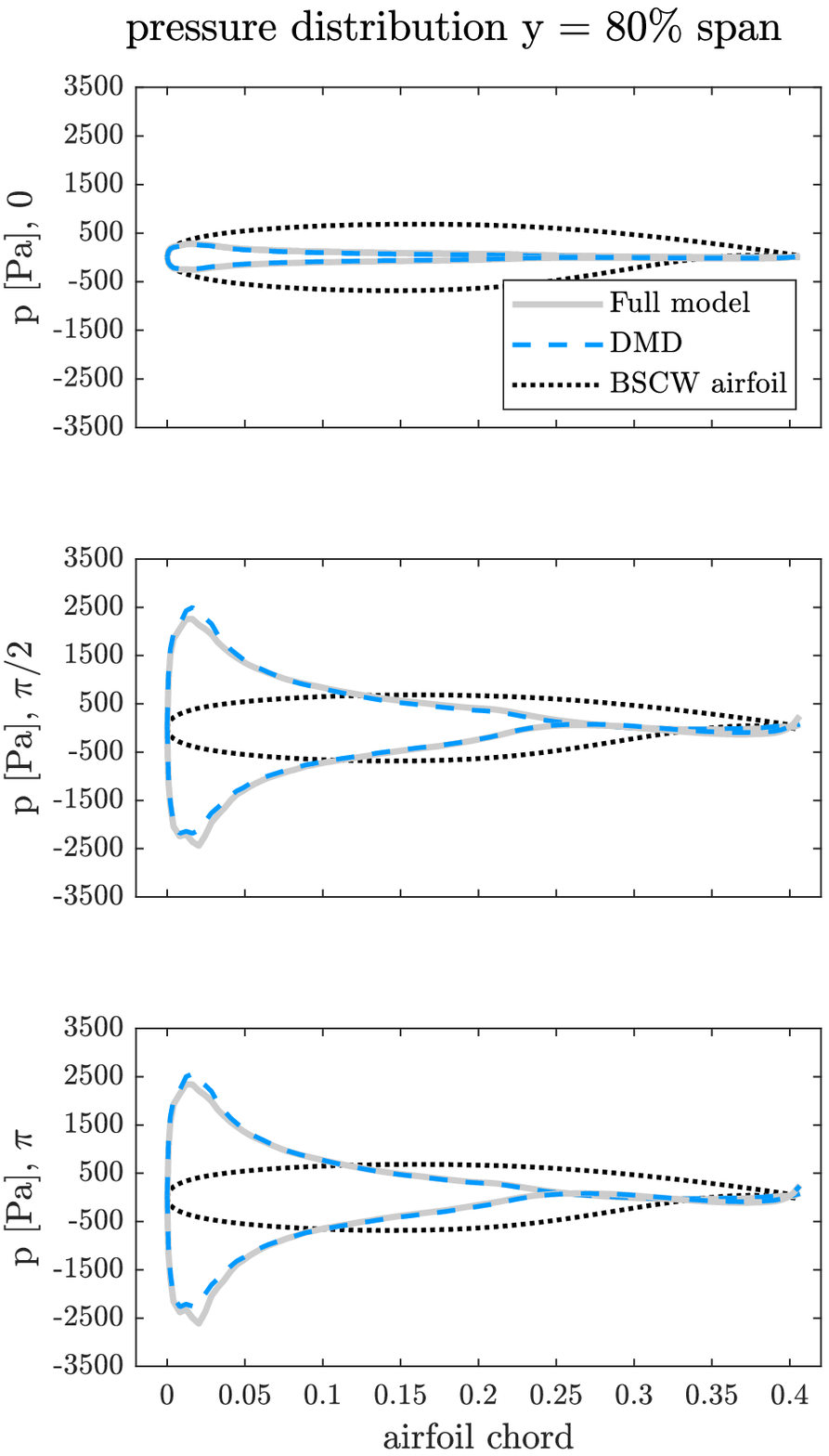}
    \caption{Pressure distribution at 20\% and 80\% of the span for the full simulation and the rank--30 DMD-ROM, around one cycle of oscillation, at the beginning of the time integration.}
    \label{fig:pressure}
\end{figure}

One important novelty of this method, with respect to previous results in the literature, is the ability to not only predict the macroscopic behaviour of the system, but also the pressure distribution on the wing. This requires the actual time integration. Predicting the wing pressure distribution is of great importance for several fields, such as control synthesis, and wing design and optimisation. For example, in Fig.~\ref{fig:pressure}, we can see the pressure distribution at 20\% and 80\% of the span for 3 time instants. The three time instants are taken at equal spacing in time, around an entire cycle. The position on the cycle is identified by the angular position as $0$, $\pi/2$, and $\pi$. 
Very good agreement is found between the full simulation and the ROM with only 30 modes. Small differences are present that explain the slight error in the damping prediction; however, the dominant physics are clearly captured.

\subsection{Sensitivity analysis}


In this section, we discuss the sensitive of our method to hyperparameters, namely, the number of modes and the amplitude of the training input signals.

\paragraph{Number of modes.}


\begin{figure}[t]
    \centering
    \includegraphics[width=\textwidth]{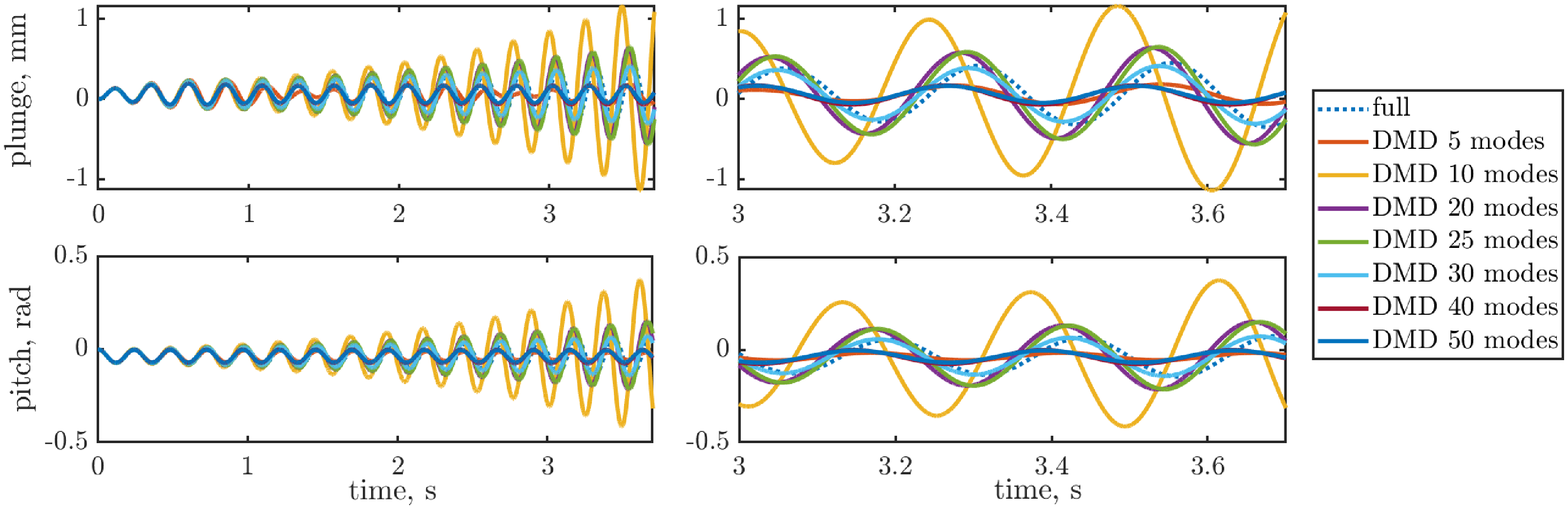}
    \caption{Pitch and plunge mode, computed with the DMD and the full simulation, for varying number of DMD modes.}
    \label{fig:differentnModes}
\end{figure}

\begin{figure}[t]
    \centering
    \includegraphics[width=0.8\textwidth]{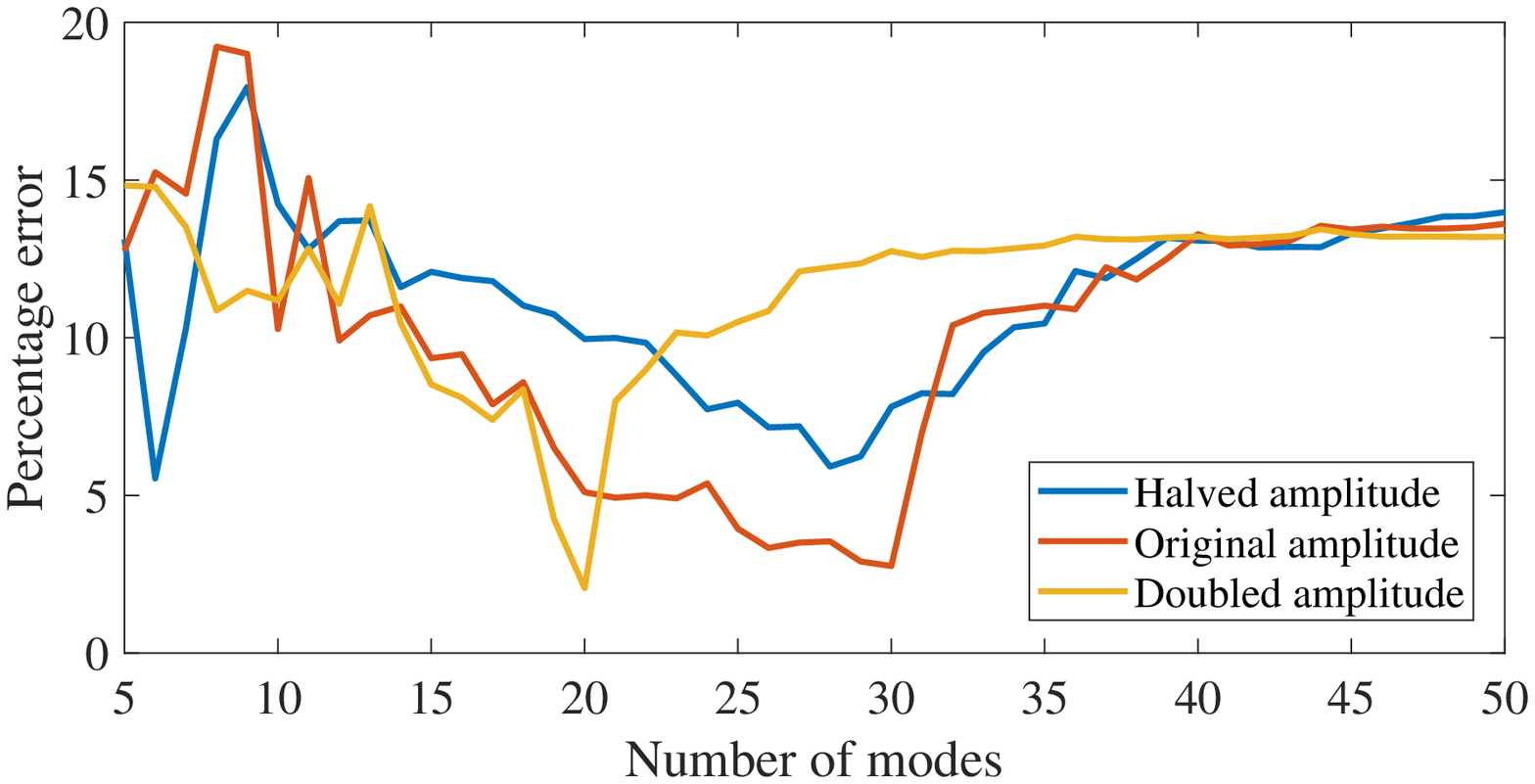}
    \caption{Errors between the DMD simulation and the full simulation of section D, for varying training amplitudes and number of modes.}
    \label{fig:error}
\end{figure}

In Fig.~\ref{fig:differentnModes}, the pitch and plunge response of the fully coupled simulation of section D is shown for different numbers of modes, namely 5, 10, 20, 25, 30, 40, and 50. The error of the ROM with respect to the full-order simulation is reported in Fig.~\ref{fig:error}. The error is computed as a norm-2 error in the structural modes trajectory. The 5 mode case is an extreme case, used to clarify how an insufficient number of modes will drastically change the system behaviour. Initially, with 10 modes, an oscillation is captured, but the frequency is incorrect. Then, the error decreases and reaches a minimum at 30 modes. Further increasing the number of modes leads to overfitting the training data, and the error increases to later flatten out. 

\paragraph{Training input signal amplitude.}

Another important parameter affecting the ROM performance is the amplitude used during training. The linearization will change depending on the chosen amplitude. With different amplitudes, different physics are exited, and different surrogate models obtained as a consequence. If the amplitude is too large, we may deviate excessively from the linearization point, while if it is insufficient, it will not excite the aerodynamics enough and the data will not contain the required amount of information.

Again, we will use the full simulation of section D for comparison. Starting from the amplitudes defined in section D, we consider halve and double the amplitude. In Fig.~\ref{fig:error}, the errors, as a function of the number of modes, are reported for the different training amplitudes. 
The three curves show a similar behaviour. For a large number of modes, they all converge to the same error that represents an overfitting of the model to the training data. Approximately the same error is also present where not enough modes are retained. In the intermediate region, a peak and a drop in the error can be observed. The peak can be explained with some included modes that must be added only in conjunction with others, not yet included, thus provoking a small divergence of the results. The drop is the optimal point we have to target.


Three important observations can be made. First, in order to find the optimal number of modes, only a single full simulation is required for training the different DMD models, thus the process is computationally efficient. Second, we can immediately observe if the provided amplitude is not sufficiently exciting the dynamics because no drop in error is seen. Third, when the amplitude is sufficiently exciting the dynamics, accurate, lower order models can be obtained with larger amplitudes. This holds true up to a certain amplitude where strong nonlinearities appear. In the context of the BSCW at the operating condition considered in this work, the appearance of strong shocks that are not present in the reference condition would indicate amplitudes too large for training accurate DMD models.

\subsection{Limitations}

\begin{figure}[t]
    \centering
    \includegraphics[width=0.47\textwidth]{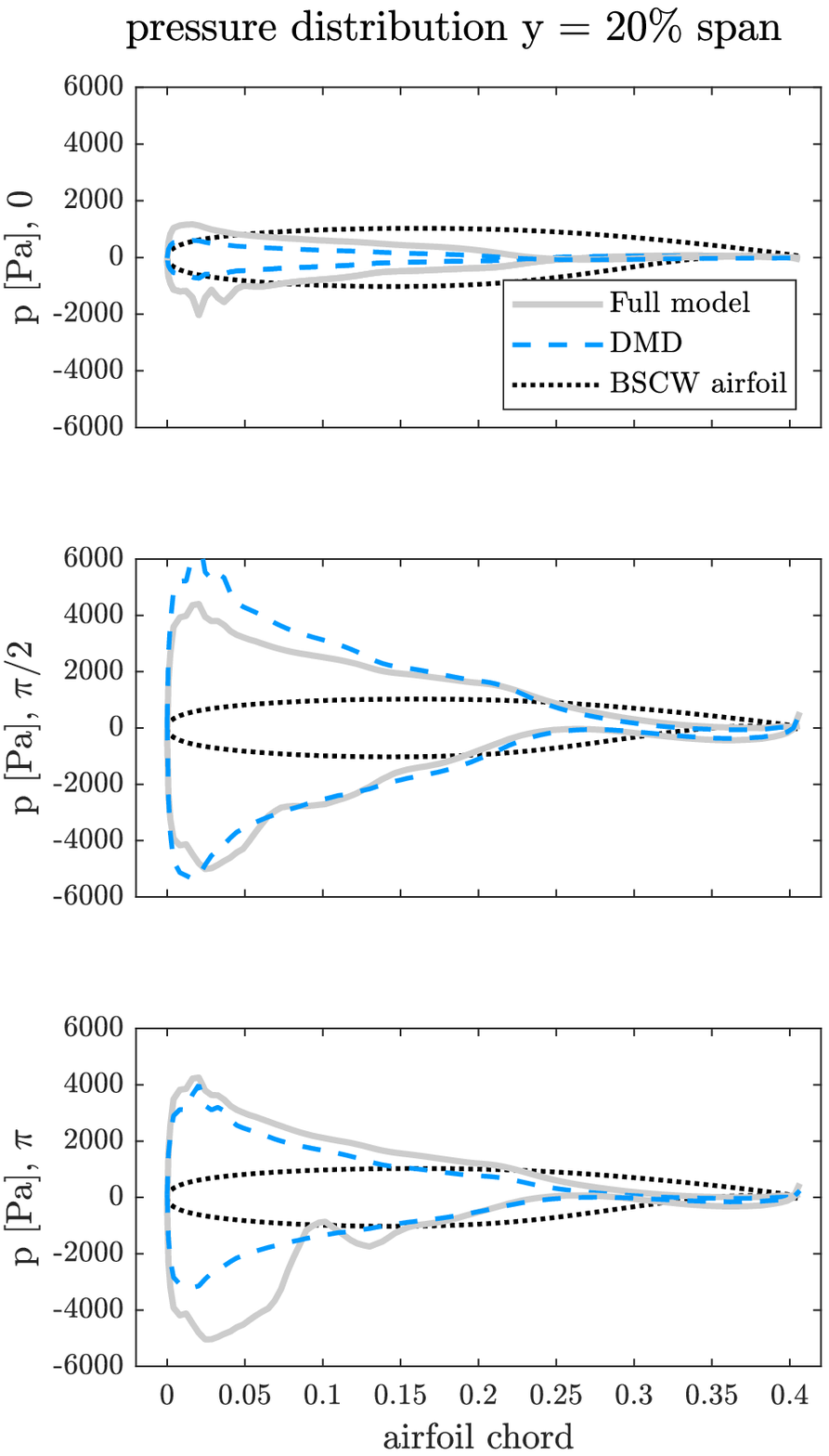}
    \hfill
    \includegraphics[width=0.47\textwidth]{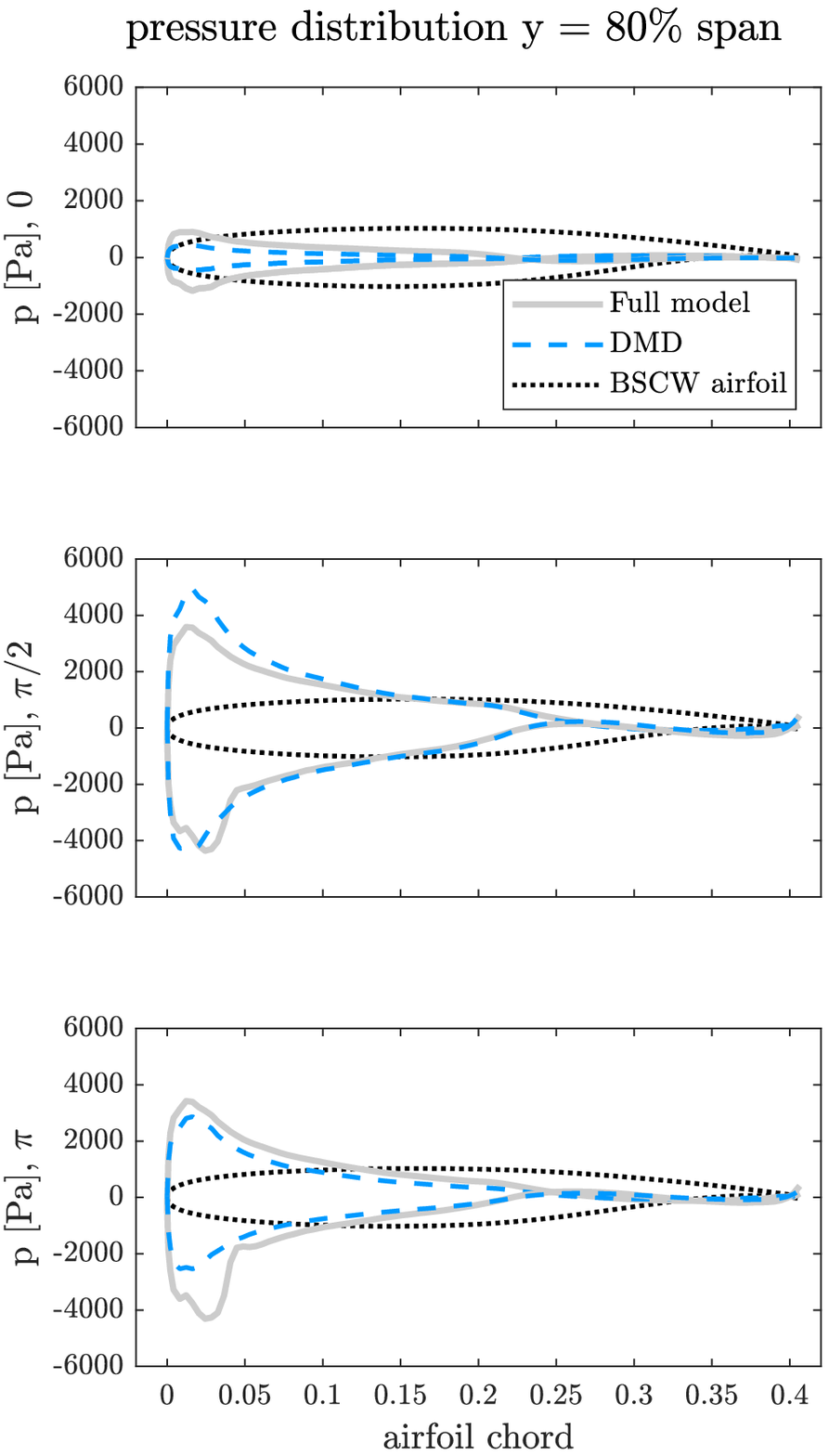}
    \caption{Pressure distribution at 20\% and 80\% of the span for the full simulation and the rank--30 DMD-ROM, around one cycle of oscillation, at the end of the time integration.}
    \label{fig:fail}
\end{figure}

The method finds a best fit \textit{linear} operator for the problem at hand, and is thus limited by the assumption of linear dynamics. 
In Fig.~\ref{fig:fail}, the pressure distributions at 20\% and 80\% of the span are reported around one cycle of oscillation at the end of simulation shown in Fig.~\ref{fig:tutorial5}. We discussed how the results of the full simulation and the DMD simulation diverge after a certain point. The model is trained around a null angle of attack where the shock wave is not present. At the end of the simulation shown in Fig.~\ref{fig:tutorial5}, large oscillations are present, particularly in the negative angle of attack range, creating a shock wave on the lower side of the wing, shown in Fig.~\ref{fig:fail} on the bottom plots. This phenomena is not present in the training data, thus, the strong shock dynamics are not captured in the DMD model. As a result, the time history of the reduced aeroelastic model diverges from the correct one. The divergence is limited by the fact that the shock wave is only present at the maximum pitch values, thus DMD still approximates the macroscopic behaviour.

Observing the pressure distribution, we can see that the reduced model is not able to represent the correct physics when the shock appears. Therefore, a second surrogate model must be built around a different linearization point where the shock is present, and the two linear models need to be interpolated. Several interpolation methods for LPV systems were recently proposed~\cite{fonzi_data-driven_2020,iannelli2021balanced,goizueta_fast_nodate}, which will be incorporated into our framework in a future step.

\section{Conclusions}

In the present work, we introduce a new method to construct aeroelastic reduced-order models from data. 
The method is based on the DMDc algorithm and is inspired by aerodynamic panel methods, which reduce the tridimensional flow around a wing problem to an equivalent bidimensional problem on the wing surface. We model the entire aerodynamics using the pressure distribution on the wing surface as a state. We use the open--source CFD solver SU2 to generate the data required for creating the ROM, and provide an open--source Python framework that can create a ROM for efficiently predicting transonic aeroelastic phenomena. A specific technique is included to stabilize the aerodynamic system in order to avoid spurious instabilities.

An important innovation introduced in this work is the ability to predict, at a significantly lower cost with respect to CFD, the pressure distribution over arbitrary 3D aerodynamic surfaces. Compared to previous research in the field of aerodynamic reduced-order modeling that focus either on full (2-D or 3-D) flow fields or on predicting global quantities, such as the lift or modal forces, here, we reduce the problem to the wing surface. This decreases the problem complexity compared to full flow field approaches and improves interpretability compared to global quantity approaches. Therefore, our method may be applied in the field of shape optimisation, morphing, and control. Specifically related to control is the ability to directly obtain an input-output relation from data, with further information on the underlying evolution of the physical state. If the structural modes include the rigid modes of the system, the B matrix includes the static control derivatives, while considering also the A matrix the dynamic derivatives can be extracted. Also, the accurate mathematical model of the entire state lends itself to observer applications and model predictive control.

In order to demonstrate the capabilities of our method, we show several test cases of increasing complexity. The final test case demonstrates that the method is able to predict the coupled aeroelastic behaviour of the BSCW wing in transonic post-flutter conditions.
The method can also be used to assess the stability of the transonic aeroelastic system in a particular condition. It is important to emphasise that a fully coupled high--fidelity simulation to verify the presence of flutter usually requires several thousand time iterations to correctly capture the damping. On the other hand, a DMD model can be trained in less than a thousand time iterations and the stability can be assessed computationally efficiently by computing the eigenvalues of the low order system. Thus, it is computationally more efficient to run a training step and build a ROM rather than assessing the stability with a time marching model, even if only one condition is examined. 
%
%
In future work, we will incorporate interpolation schemes for surrogate models at different operating conditions~\cite{Hemati2016aiaa,fonzi_data-driven_2020,iannelli2021balanced}, which will extend the present method to be used as a flutter search tool.

\section*{Funding Sources}
The authors acknowledge support from the Air Force Office of Scientific Research (AFOSR FA9550-19-1-0386)
and the National Science Foundation AI Institute in Dynamic Systems (Grant No. 2112085).

\section*{Acknowledgments}
Code: \href{https://github.com/Nicola-Fonzi/pysu2DMD}{https://github.com/Nicola-Fonzi/pysu2DMD}

 \begin{spacing}{.9}
 \small{
 \setlength{\bibsep}{6.5pt}
 \bibliographystyle{RS}
 \bibliography{biblio}
 }
 \end{spacing}

\end{document}